\documentclass{anonymous_2}
\usepackage{natbib}\usepackage{txfonts}\usepackage{balance}
\usepackage{graphicx}
\idline{75}{1}
\begin{document}
\def\teff{$T\rm_{eff }$}
\def\kms{$\mathrm {km s}^{-1}$}

\title{
The Chromosphere During Solar Flares
}

   \subtitle{}

\author{
L. \,Fletcher%\inst{1} 
 }

  \offprints{L. Fletcher}

\institute{
Department of Physics and Astronomy,
University of Glasgow,
Glasgow G12 8QQ,
United Kingdom
\email{lyndsay@astro.gla.ac.uk}
}

\authorrunning{Fletcher}

\titlerunning{Flare Chromospheres}

\abstract{The emphasis of observational and theoretical flare studies in the last decade or two has been on the flare corona, and attention has shifted substantially away from the flare's chromospheric aspects. However, although the pre-flare energy is stored in the corona, the radiative flare is primarily a chromospheric phenomenon, and its chromospheric emission presents a wealth of diagnostics for the thermal and non-thermal components of the flare. I will here review the chromospheric signatures of flare energy release and the problems thrown up by the application of these diagnostics in the context of the standard flare model. I will present some ideas about the transport of energy to the chromosphere by other means, and calculations of the electron acceleration that one might expect in one such model.
}
\maketitle{}

\section{Introduction}
The first recorded observation of a solar flare was made on 1st September 1859 \citep{1859MNRAs..20...13C} and showed features which we would now recognise as broadly chromospheric in origin. The two small `patches of intensely bright and white light', now termed a white-light flare, mapped the locations from which the bulk of the solar flare energy was radiated. Although Carrington could not have known it at the time, the flare was accompanied also by the emission of ultraviolet to X-ray and most probably also $\gamma$-ray radiation. These radiations, also primarily chromospheric in origin, produced changes in the ionospheric ionisation level, leading to currents disturbances and strong geomagnetic activity, recorded as an abrupt magnetometer deflection at Kew Gardens in London. 

For decades, studies of solar flares were made primarily via their lower atmospheric, and particularly their chromospheric signatures. Due to its low contrast with the surrounding photosphere,  white-light flares  are hard to observe, but the flare dynamics and evolution are shown wonderfully well by the bright H$\alpha$ emission, usually organised into two or more elongated `ribbons'. Observations of these H$\alpha$ ribbons, as well as the motions and disappearance of the
H$\alpha$ filaments and the post-flare growth of H$\alpha$ loops was in great part responsible for the development of  the flare `standard flare model'.

As solar physics instrumentation moved into space, much of the attention of the solar flare community shifted to imaging and spectral observations at higher energies: the ultraviolet, extreme UV,  X-ray and $\gamma$-ray. In particular, the spectacular EUV and soft X-ray loops of the reconfiguring coronal plasma, with their close link to the all-important coronal magnetic field, have been very compelling. But energetically the EUV and X-ray emissions are a sideshow \citep{2004JGRA..10910104E,2005JGRA..11011103E}. The optical to ultraviolet continuum and lines are the primary means by which the solar atmosphere rids itself of energy during a flare. To understand solar flares, we must link what we have learned of coronal processes during the last decades of space-based observation with what the chromosphere has to tell us.

\begin{figure*}[t!]\label{fig:fig1}
\includegraphics[width=\textwidth]{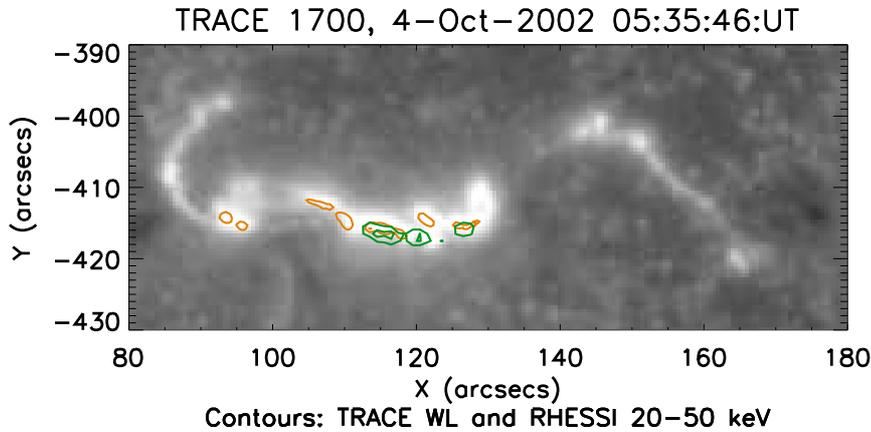}
\caption{\footnotesize A composite image showing the GOES M8.5 flare on 04-October-2002. The background image is the TRACE~1700~\AA\ channel (log scaled), on which are superposed contours showing the strongest white light patches (yellow/light) and the RHESSI 25-50 keV sources, made using the Pixons technique. The relative pointing between TRACE and RHESSI is uncertain, so the RHESSI sources have been translated by (-5, 1) arcseconds to obtain agreement with the strongest white-light sources. }
\end{figure*}

\section{Flare chromospheric sources}
The impulsive phase flare morphology  in H$\alpha$, UV and chromospheric EUV is characterised by elongated ribbons of emission which start out close together, on either side of a magnetic neutral line, and lengthen and spread outwards with time. Within these ribbons there are compact, bright sources  (kernels) which tend to be located on the outer edges of the ribbons. Overlaying images made in hard X-rays (HXRs) with the RHESSI mission \citep[e.g.]{2007ApJ...654..665T} shows clearly that kernels are co-spatial with HXR chromospheric sources, generally known as flare `footpoint' sources. Only a small number of HXR footpoint sources are visible in a flare, probably due in part to the limited dynamic range offered by the indirect imaging techniques in the HXR part of the spectrum. However the very fact that there are sources with significantly higher photon fluxes points to the existence of special locations within the overall flare geometry. It is also not surprising that these locations are where the strongest enhancements in white light are to be found. This has been shown using TRACE observations \citep{2003ApJ...595..483M,2006SoPh..234...79H,2007ApJ...656.1187F} taken through the `white light' (WL) open filter. In the case of this filter, `WL' may be a misnomer due to the high flux of UV radiation that it also admits. However, emission in TRACE WL and in UV shows very different spatial characteristics, with ribbons in UV that are extended compared to the concentrated, compact WL sources where most of the flare energy enters the chromosphere. Figure~\ref{fig:fig1} shows the spatial relationship between UV flare ribbons, WL sources and HXR sources.

The locations of strong flare chromospheric emission correspond to locations of particular significance in the overall magnetic structure of the flare. The standard `CSHKP' model links ribbons, flare loops, and the supposed coronal magnetic reconnection site, in a simple 2-D geometry. This has been transported into 3D and put on a firmer theoretical footing, starting with the work of \cite{1992SoPh..139..105D} who found correspondence between the photospheric projections of strong quasi-separatrix layers, and the chromospheric H$\alpha$ sources. Subsequent work, for example by \cite{2003ApJ...595..483M} has confirmed this, and the most recent thinking in theory \citep[e.g.]{2009ApJ...693.1628J} associates the HXR and by implication the WL sources with the lower atmosphere mapping of singular field lines known as spines, which join two magnetic nulls. Reconnection at such structures is often proposed to lead to flare particle acceleration, but to provide the accelerated electrons required to explain the observed HXR radiation in the collisional thick target model (CTTM) needs a volume of flaring corona outlined roughly by the extent of the flare ribbons and the flare loops to be `processed' by the accelerator each second (see Section~\ref{sect:energetics}).  It is unlikely that acceleration can take place in a current sheet, null, separator or QSL, because the throughput of plasma through these singular structures is - even assuming 100\% acceleration efficiency and a high Alfv\'en speed - just too small for reasonable impulsive phase source dimensions. The electron requirements point towards a truly volumetric acceleration mechanism operating throughout a large portion of the corona local to the flare. But then how can all the electrons accelerated there be directed towards the footpoints? Electrons follow the magnetic field, and the field permeating the flaring corona certainly is not all rooted in the flare footpoints.  

\section{Chromospheric Dynamics}

The response of the chromosphere to the energy deposited there during a solar flare is to radiate, conduct and expand - the latter is usually  known as chromospheric `evaporation' although it has nothing to do with the usual meaning of the word. Evaporation can happen gradually or explosively, depending on whether the heating timescale is larger or smaller than the hydrodynamic expansion time of the chromosphere. Evaporation has been observed using spectroscopy, originally with $H\alpha$ and unresolved SXR line profiles \citep{1988ApJ...329..456Z} and latterly in the EUV with imaging spectrometers.  In the few flare impulsive footpoints which have been observed using EUV spectroscopy, the usual pattern is that higher temperature lines show a blueshift from the upflowing plasma, and lower temperature lines are redshifted. The redshift is interpreted as  due to a downwards-moving `chromospheric condensation', driven by a pressure pulse which occurs when the upper chromosphere is heated rapidly past the point where it can radiate efficiently, and expands quickly. There is rough momentum balance between these two components.

However, with  {\it Hinode}/EIS, \cite{2008ApJ...680L.157M} and  \cite{2009ApJ...699..968M} have recently observed redshifted emission components up to 1.5~MK - much higher temperatures for downward-moving plasma than is  expected from theoretical calculations. This would require heating of the upper chromosphere, which could be delivered - in accordance with the flare coronal electron beam (Section~\ref{sect:energetics}) by a soft spectrum of electrons in a beam from the corona, such as is suggested by the large electron spectral index deduced by \cite{2009ApJ...699..968M}.  But  most mysteriously, at high energies \cite{2009ApJ...699..968M} also find dominant, {\it stationary} high temperature ($\ga 12$~MK) footpoint sources in the impulsive phase, which are not expected in a beam heating model at all.

\section{Flare chromosphere energetics}~\label{sect:energetics}
The chromosphere is highly structured in temperature and density, horizontally as well as vertically, and also changes from being high-$\beta$ to low-$\beta$, from almost neutral to fully ionised, and from optically thick to optically thin - at all wavelengths - over a matter of a few thousand kilometers. It is a complicated enough matter to interpret chromospheric radiation in the nominally `quiet' chromosphere. In a flare, the chromospheric energetics is entirely dominated by the flare energy release, which accounts for a power per unit area of around $10^{11} {\rm erg\;cm^{-2}s^{-1}}$ being dumped in the chromosphere (this number may increase, as we have not yet resolved chromospheric X-ray and optical/UV sources). At least during the impulsive phase, with such a dominant and rapid heating, and interruption to the normal chromospheric state of affairs, the normal detailed chromospheric structuring may become irrelevant, and we should think instead of a fairly homogeneous plasma, which rapidly heats, ionises, radiates and expands - a chromospheric `fireball' (terminology due to H. Hudson).

The origin, transport and conversion of the energy powering the chromospheric flare is a primary question in solar physics. It is indisputably the case that the flare energy is stored in the stressed magnetic fields of the pre-flare corona. In the standard flare model electrons (and possibly also ions), accelerated in the corona and channeled by the magnetic field into the chromosphere are the primary agents that transport the energy to the chromosphere, where it is radiated. Since the 1970s, the beam model and the associated `collisional thick target model'  (CTTM)  interpretation of HXR radiation has been widely accepted\citep{1970SoPh...13..471S,1971SoPh...18..489B,1972SoPh...24..414H}. The bremsstrahlung emission from electrons propagating in the dense chromosphere, though itself energetically insignificant, is a major diagnostic of the flare electron total energy, spectral and spatial distribution, and also more recently their angular distribution \citep{2006ApJ...653L.149K}. Within the framework of the CTTM it has been possible to deduce the energy and number fluxes of electrons arriving from the corona at the chromosphere. However, here a possible problem emerges, in the deduction that the number of electrons required per second to power the bremsstrahlung emission inferred from the CTTM is large compared to the number of electrons available in a reasonable coronal volume and, more seriously, that the return current generated by such a beam travels so fast in the corona that it is - according to our current understanding -  unstable \citep{1977SoPh...52..117B,2003LNP...612...58F}. The effect of an instability is not clear, but it is bound to extract energy from the propagating electron distribution resulting in heating as it propagates \citep{1988SoPh..115..289C,2003JGRA..108.1442P} and may halt it altogether. If the return current drift speed exceeds the sound speed, the result is the onset of ion acoustic turbulence, as discussed by \cite{1976SoPh...48..197H}, while the Buneman instability results if the return current drift speed exceeds the electron thermal speed \citep{1959PhRv..115..503B}. Under certain circumstances, the Buneman instability can, on saturating, itself accelerate electrons \citep[e.g][]{2003Sci...299..873D} but the initial energy converted to heating has already sapped the beam energy.

Stability depends on the beam number flux rate (number per second per unit area). The question of stability against ion acoustic turbulence was raised with early observations \citep{1976SoPh...48..197H} and now with improving observational techniques the observed small sizes of HXR sources means that the Buneman instability also starts to be of concern. For example,  collisional thick-target interpretation of the 23 July 2003 flare implies a few (up to around 5) $\times 10^{36}{\rm electrons\;s}^{-1}$ at the chromospheric HXR source location at the peak of the flare \citep{2003ApJ...595L..97H}. During flare maximum, there are three main HXR chromospheric sources, each around 5"$\times$5" giving a reasonable value for the beam area in the chromosphere of $4 \times 10^{17}{\rm cm^{-2}}$  (the source size may in fact reflect the RHESSI angular resolution, and white light footpoints also suggest smaller sources than this). Thus an electron flux per unit area of around $10^{19}{\rm cm}^{-2}{\rm s}^{-1}$ is required. The low energy cutoff at this time is 20~keV, and the power law index around 6, so the bulk of the electrons have energies from 20-40~keV and are traveling at $8.4 - 11.9 \times 10^9 {\rm cm\;s}^{-1}$ - we take $v_{\rm beam} = 10^{10}{\rm cm\;s}^{-1}$ as representative. Thus the beam density at the location in the chromosphere where the HXRs are generated is around $n_{\rm beam} = 10^{10}{\rm cm}^{-3}$. Even for the upper chromosphere this is not a trivial density perturbation and,  ignoring for now the magnetic field convergence between corona and HXR source location,  it implies that the coronal beam density is a substantial fraction, or even in excess of, expected coronal densities. Typically $n_{\rm cor}$ is a few $\times 10^9{\rm cm}^{-3}$ early in a flare, increasing to a few times $10^{11}{\rm cm}^{-3}$, once evaporation starts. The return current speed. $v_{rc}$, given by $n_{beam} v_{beam} = n_{cor} v_{rc}$ is therefore a substantial fraction of the beam speed - vastly in excess of the ion sound speed, and greater than the electron thermal speed which at 20~MK is $3\times 10^9{\rm cm\;s}^{-1}$. For a given total electron rate the density of beamed electrons in the corona can be reduced if one takes into account magnetic field convergence - the beam density varies inversely with the magnetic mirror ratio. But then magnetic mirroring results in a higher overall number of non-thermal electrons (i.e. in a non-beam distribution), including those trapped in the corona. The beam can travel stably if the density of the loop in which it moves is large (see e.g. \cite{1990A&A...234..496V} for details) but though this is possible later in the flare, once evaporation has started, there is no imaging or spectroscopic evidence before the flare for such loop densities (except in coronal thick target loop flares \citep{2004ApJ...603L.117V}, which do not show footpoints). Invoking evaporative resupply also precludes any model involving coronal acceleration onto loops which are connected with a reconnection region, since once the evaporation starts the loops have long since detached from the accelerator. So, it is clear that there are reasons for seeking alternatives to this standard beam model.

\section{Chromospheric  acceleration in solar flares}
In view of the difficulties with supplying and stabilising a coronal electron beam with the flux required to explain collisional thick target hard X-rays, new models have recently emerged for solar flare energy transport and electron acceleration, with the chromosphere at their heart. \cite{2008ApJ...675.1645F} investigated a model in which the stored coronal magnetic energy is transmitted as an Alfv\'enic 'pulse' to the chromosphere, there to be dissipated, resulting in chromospheric heating and electron acceleration. \cite{2009A&A...508..993B} have suggested that electrons are accelerated, or re-accelerated, locally in the chromosphere by parallel electric fields in many small-scale current sheets. The two ideas have several aspects in common - the requirement to develop small-scale structure in the chromosphere, the need to impart magnetic stress to the chromosphere (i.e. to generate the current sheets in the Brown et al. model) and of course the notion of tapping into the chromospheric electron reservoir to supply the electrons that produce the X-rays. It should be highlighted here that evidence for coronal electron acceleration most certainly exists, in the form of non-thermal coronal HXR sources \citep{2008A&ARv..16..155K}, radio Type III bursts \citep[e.g.]{1995ApJ...455..347A}, and the energy-dependent HXR delays \citep{1995ApJ...447..923A} interpreted as faster electrons accelerated in the corona arriving at a chromospheric target before slower ones. However, the demands on the electron number implied by all of these signatures are modest. Although the coronal HXR sources may in some cases require that a large fraction of coronal electrons in the source are accelerated, magnetic trapping and a long coronal collision time means that there is no resupply problem. Type III bursts are non-linear plasma radiation, and are thought not to require substantial electron numbers. And the HXR bursts from which the energy-dependent delays are deduced are typically small (rarely more than 20\% total HXR fluence) on top of a more slowly varying background \citep{1996ApJ...470.1198A} which has an opposite dependence of delay on energy \citep{1997ApJ...487..936A}. So, while electron acceleration in the corona is almost certainly taking place (and indeed fits well within the framework of the \cite{2008ApJ...675.1645F} model), the fact remains that it is difficult to see how this can provide also the chromospheric electrons.

\begin{figure*}[t!]\label{fig:fig2}
\hbox{
{\includegraphics[width=0.5\textwidth]{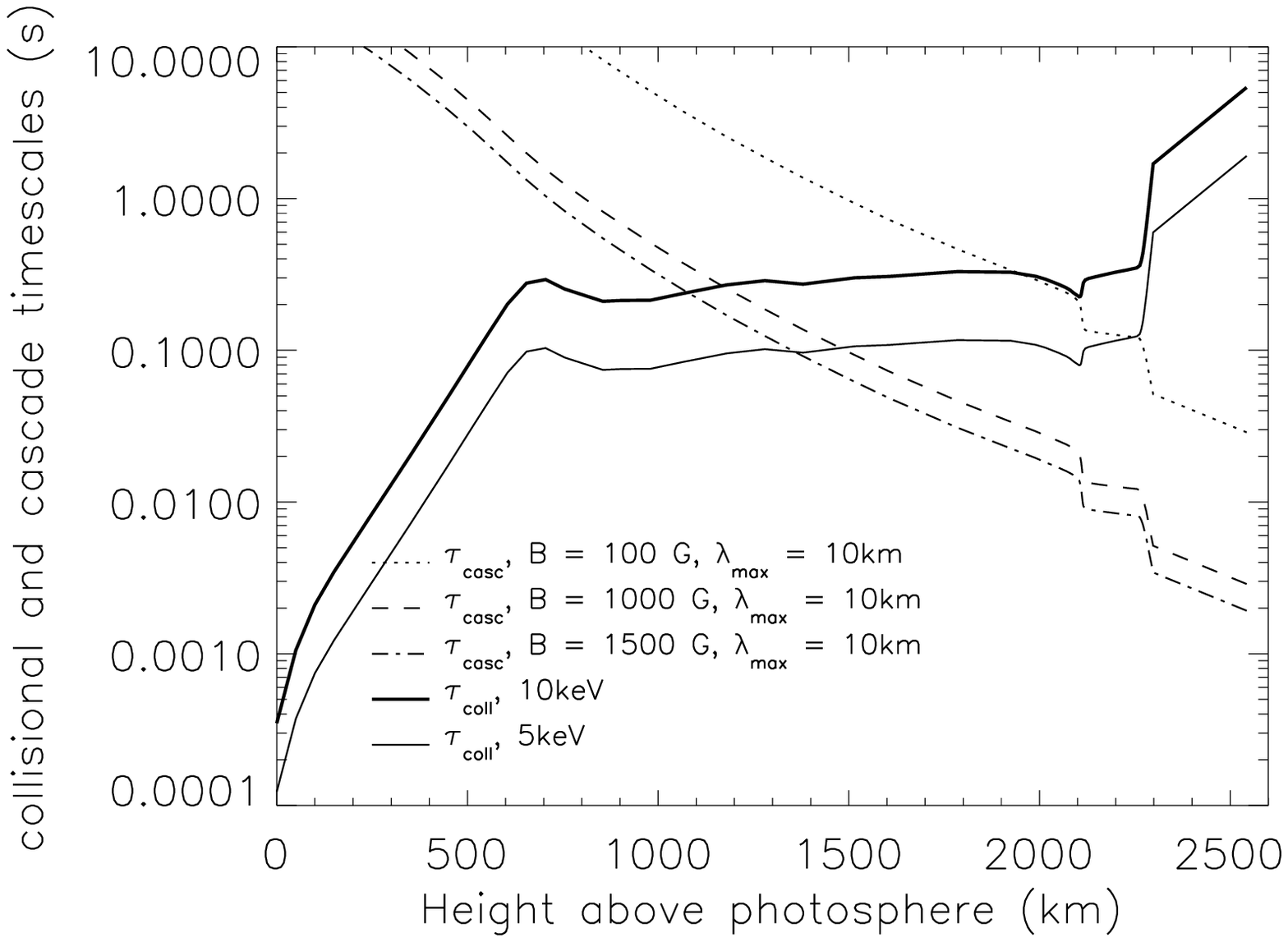}}
{\includegraphics[width=0.5\textwidth]{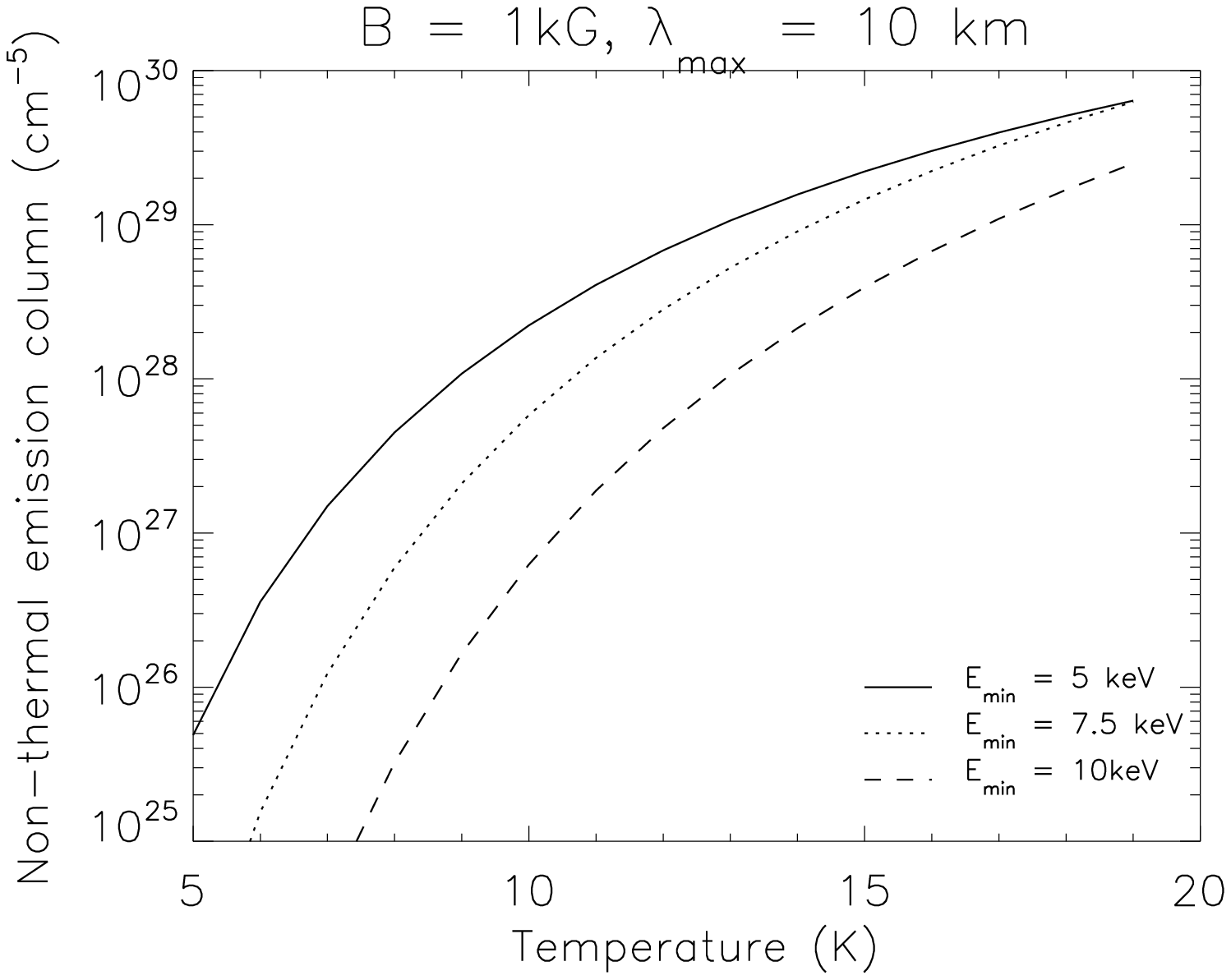}}
}
\caption{\footnotesize {\it LH Panel}: the collisional timescales based on electron-ion scattering (solid lines), and the cascade timescales (dashed lines) calculated for the VAL-C model atmosphere parameters. RH Panel: The non-thermal column available in the accelerating volume as a function of its temperature, at different values of the energy at which a particle is considered 'non-thermal' and collisionless. }
\end{figure*}

Chromospheric electron acceleration, as opposed to heating, is feasible only under certain circumstances, as the chromosphere is dense and highly collisional. Even without knowing in detail the physics of the possible acceleration mechanisms some general statements can be made. A mechanism which energises the chromospheric electrons in an isotropic manner will first result in heating, as the electron-electron collisional timescale is short, and the electron component will rapidly adopt a Maxwellian distribution.  This timescale is given by $\tau_{ee}=3.4\times 10^{5}{T_e^{3/2} /n_e\Lambda_{ee}}$ seconds, where $T_e$ is expressed in eV and $\Lambda_{ee}$ is the electron-electron Coulomb logarithm (see NRL Plasma Formulary). In a chromospheric density of $10^{11}{\rm cm}^{-3}$ , Coulomb logarithm $\Lambda_{ee} \sim 20$ with a pre-flare chromospheric temperature of $10^4$~K or 1eV, we have  $\tau_{ee} = 1.7\times 10^{-7}$~s. As the electron distribution heats the timescale will increase. At $\sim 1$~MK  $\tau_{ee} = 1.7\times 10^{-4}$~s. 

If the acceleration mechanism results in an anisotropic electron distribution (e.g. as would be the case for field-aligned current sheets or from the electric fields generated by kinetic or inertial Alfv\'en waves) then the physics will be slightly different, as a drifting Maxwellian will result, with all electrons receiving an additional large velocity boost in the same direction (and all ions in the opposite direction). Then the more relevant timescale for thermalisation of the electron population is that on which the electrons scatter on ions, which also have an oppositely drifting Maxwellian distribution. This scattering renders the electron distribution once more isotropic and able to relax collisionally to a hotter Maxwellian. This timescale, in the case of the kinetic energy $\epsilon$ of the accelerated particles being substantially greater than the temperature of the Maxwellian, is $\tau_{\perp,ei} = 1.3\times 10^5 \epsilon^{1/2}/n_i\Lambda_{ei}$ where $\Lambda_{ei}$ is the electron-ion Coulomb logarithm. This is also a short timescale. 

It is well-known that the chromosphere heats substantially during the impulsive phase, for example the footpoint stationary components in highly-ionised lines of iron observed by {\it Hinode}/EIS \citep{2009ApJ...699..968M} up to $ T_e$ = 16~MK (assuming ionisation equilibrium, which is an assumption requiring examination in this context) and possibly higher \citep{2009ApJ...699..968M}. Also {\it Yohkoh}/SXT observations of so-called `impulsive soft X-ray footpoints' \cite{2004A&A...415..377M} show footpoint temperatures averaging 8-10~MK and densities from $4 - 10\times 10^{10}\rm{cm}^{-3}$. As the footpoint temperature increases, an increasing fraction of the electron distribution becomes effectively collisionless. For example,  at 10~MK,  close to 1\% of electrons have energy $\epsilon$ above 5~keV (at 15~MK this is 5\%).  If we assume anisotropic acceleration in a plasma of density $10^{11}\rm{cm}^{-3}$, the electron-ion, and the electron-electron relaxation timescales for electrons above this energy is $\sim$ 0.02s.  Thus, an accelerator operating on a shorter timescale will accelerate these electrons further. 

As an example, consider acceleration by a wave-based mechanism, starting with the arrival of a macroscopic MHD perturbation at the top of the chromosphere, which cascades to shorter spatial scales where energy can be picked up by particles. As in the case of wave-particle acceleration in the corona, the longest timescale in the system is the turnover time of the largest perpendicular wavelength, $\tau_{casc} = (\lambda_{max}/v_A)(B/\delta B)$ which we take as an upper limit to the acceleration timescale. The perpendicular wavelength is important since the chromospheric magnetic field is so strong that a fully 3D cascade will not happen. Figure~\ref{fig:fig1} shows the perpendicular cascade timescale and collisional timescale (using electron-ion collisions) calculated in the VAL-C model \cite{1981ApJS...45..635V}, plotted for a perpendicular wavelength of 10~km, various values of the chromospheric magnetic field strength, and a field perturbation of 5\%. Note, in the VAL-C model the chromospheric electron density varies between $2 \times 10^{10}{\rm cm^{-3}}$ at  2200km and $6\times 10^{10}{\rm cm^{-3}}$ at 1600~km. The figure shows that the acceleration timescale calculated in this way is smaller than the thermalisation timescale throughout the top few hundred km of chromosphere.

So substantial chromospheric acceleration is clearly a possibility. In the `local reacceleration' scenario of \citep{2009A&A...508..993B}, the energy gain by particles in a chromospheric acceleration volume offsets the collisional losses that they experience there, but the electrons could also escape the acceleration volume and radiate in the surrounding collisional thick target as normal, with the usual attendant requirement on replenishment of the accelerator. The total requirement for the non-thermal emission measure of a flare \citep{2009A&A...508..993B} is $f n_h n_e V \sim 10^{46}$, where $f$ is the fraction of all electrons in volume $V$ that are accelerated, and $n_h$, $n_e$ are the hydrogen and electron number densities \cite{2009A&A...508..993B}. We note that the Bremsstrahlung cross-section is almost the same for neutral and for ionised hydrogen, which is why $n_h$ is used rather than the proton number density. Looking first at the case that the acceleration and the radiation volume are one and the same, and taking the lower boundary of the acceleration volume as the depth at which the thermalisation timescale at 5~keV equals the cascade timescale, results in a column-integrated value $\int n_h \times n_e dz \sim 4\times 10^{30}{\rm cm}^{-5}$.   For $f = 0.01$ throughout most of this volume, to match the required non-thermal emission measure implies that a source area of  $2.5 \times 10^{17}{\rm cm^{2}}$ is required.  Varying the chromospheric plasma temperature results in different fractions of electrons in the high-energy tail, and different requirements on the flare area. The right-hand panel of Figure~\ref{fig:fig2}, shows the non-thermal column $f n_h n_e V dz$ as a function of electron temperature, at different values of the minimum energy above which the number of particles in the Maxwellian tail is evaluated. For example, if the electron temperature is 15~MK and assuming a magnetic field strength of 1~kG, and magnetic perturbation as before, a non-thermal column of $2\times 10^{29}{\rm cm}^{-3}$ can be produced, requiring a flare area of $5\times 10^{16}{\rm cm}^2$ to produce the overall non-thermal emission measure.

Accelerated electrons can also escape the acceleration volume and enter denser regions lower down, where they undergo collisional thick target radiation as usual, without further acceleration. In this case the non-thermal emission measure would be $N_{\rm stop} f n_e A$ where $N_{\rm stop}$ is the collisional stopping depth, $f n_e$ the density of fast particles from above and $A$ the flare area. For a 30~keV electron, the collisional stopping depth in a fully ionised plasma is $\sim 3 \times 10^{20}{\rm cm}^{-2}$, and somewhat larger in a partially-ionised plasma because of the smaller effective Coulomb logarithm. In the VAL-C model, the location of most of the electron acceleration is at a density of around $4\times 10^{10}{\rm cm}^{-3}$, so that $f n_e = 4\times 10^{8}{\rm cm}^{-3}$, and the non-thermal emission measure would be $\sim 10^{29}~A~{\rm cm}^{-3}$.  So again, an area of $10^{17}{\rm cm}^2$ would provide the necessary radiation. The stability of electron re-supply in the form of a return current is more likely because of the low value of the accelerated fraction $f$. However, it should be noted that this scenario is not necessarily consistent with the results of \cite{2006ApJ...653L.149K} who find evidence for a basically isotropic distribution of radiating electrons. It remains to be seen whether the combination of electron scattering and mirroring could generate such a distribution, though we note that in the lower chromosphere, where the number of free electrons is reduced compared to the number of ions/hydrogens, that electron scattering will happen more rapidly than electron energy loss (c.f. \cite{1978ApJ...224..241E}. Eq. 30), which is a favourable situation for producing an isotropic but still energetic distribution.

\section{Conclusions}
This brief summary sketches out the observational basics of chromospheric flares, though ignoring various aspects such as what is know from optical and UV spectroscopy (see article by Hudson et al. in this volume), and $\gamma$-ray radiation.It outlines how the normal model of flare energy transport by a beam of electrons from the corona may run into difficulty, and discusses alternatives to this model which invoke chromospheric electron acceleration. Such a model appears to be very feasible, assuming that the chromospheric plasma can also be heated as part of the process - wave- or current-driven - that transports energy to the chromosphere. This article has not touched at all on the issue of generating the flare white light signature, which is where the chromospheric flare story started, but since the problem with existing models has always been how to get sufficient energy to a sufficient depth in the chromosphere using an electron beam accelerated in the corona, it is clear that chromospheric acceleration models offer potential solutions to this problem, as was pointed out also by \cite{2009A&A...508..993B}.

\begin{acknowledgements}
The author is grateful for travel support from the NSO, which enabled her to attend the very stimulating workshop at which this presentation was made. This work is also supported by the EUÕs SOLAIRE Research and Training Network at the
University of Glasgow (MTRN-CT-2006-035484) and by Rolling Grant ST/F002637/1 from
the UKÕs Science and Technology Facilities Council.
\end{acknowledgements}

\bibliographystyle{aa}
\bibliography{fletcher_nso}

%\begin{thebibliography}{}

%\bibitem[\protect\citeauthoryear{Milligan 
%\& Dennis}{2009}]{2009ApJ...699..968M} Milligan R.~O., Dennis B.~R., 2009, ApJ, 699, 968 

%\bibitem[\protect\citeauthoryear{Liu, Petrosian, 
%\& Mariska}{2009}]{2009ApJ...702.1553L} Liu W., Petrosian V., Mariska J.~T., 2009, ApJ, 702, 1553 

%\end{thebibliography}

\end{document}